# A PRELIMINARY STUDY OF THE LINEAR RELATIONSHIP BETWEEN MONTHLY AVERAGED DAILY SOLAR RADIATION AND DAILY THERMAL AMPLITUDE IN THE NORTH OF BUENOS AIRES PROVINCE

# RELACIÓN LINEAL ENTRE LOS PROMEDIOS MENSUALES DE RADIACIÓN SOLAR Y AMPLITUD TÉRMICA EN SAN NICOLÁS, NORTE DE LA PROVINCIA DE BUENOS AIRES

**R. Cionco, N. Quaranta[1], R. Rodriguez**
Universidad Tecnológica Nacional-Facultad Regional San Nicolás, Colón 332, San Nicolás, c.p. 2900, Buenos Aires
Tel.: +54 -336- 4420830; fax: +54- 0336 4425266 e-mail: gcionco@frsn.utn.edu.ar
[1] Investigador CIC-PBA

**ABSTRACT**

Using irradiance and temperature measurements obtained at the Facultad Regional San Nicolás of UTN, we performed a preliminary study of the linear relationship between monthly averaged daily solar radiation and daily thermal amplitude. The results show a very satisfactory adjustment (R = 0.848, RMS = 0.066, RMS% = 9.690 %), even taking into account the limited number of months (36). Thus, we have a formula of predictive nature, capable of estimating mean monthly solar radiation for various applications. We expect to have new data sets to expand and improve the statistical significance of these results.

**RESUMEN:** Utilizando medidas de radiación solar global y temperatura obtenidas en la Facultad Regional San Nicolás de UTN, se realizó un estudio preliminar de la relación lineal entre los promedios mensuales de radiación solar diaria y las medias mensuales de amplitud térmica diaria. Los resultados muestran un ajuste muy satisfactorio (R = 0,848; RMS = 0,066; RMS% = 9,690 %), aun teniendo en cuenta el número limitado de meses (36). De esta forma, se dispone de una fórmula de carácter predictivo, capaz de estimar medias mensuales de radiación solar para aplicaciones diversas. Se espera disponer de nuevas series de datos para ampliar y mejorar la significación estadística de estos resultados.

**Palabras clave:** radiación solar diaria, predicción, amplitud térmica, norte de Buenos Aires.

## INTRODUCCIÓN

La Universidad Tecnológica Nacional a través del Grupo de Estudios Ambientales de la Facultad Regional San Nicolás, realiza desde hace algunos años un trabajo de monitoreo de calidad de aire mediante convenios específicos con instituciones públicas y privadas; en particular, con las principales empresas del importante cinturón industrial de la zona. Como parte de esta actividad, se obtienen datos meteorológicos con diversas estaciones automatizadas, una de las cuales, ha realizado mediciones de irradiancia solar. Esos datos han comenzado a ser liberados para usos diversos, lo cual motivó el presente trabajo.

El norte de la provincia de Buenos Aires, especialmente la zona de San Nicolás de los Arroyos ($\varphi = -33°\,19'$, $\lambda = -60°\,12'$) es una de las regiones de mayor actividad industrial y agrícola del país, además de presentar gran variabilidad climática asociada a fenómenos como El Niño (Rusticucci y Vargas, 2002). La medición directa de la radiación solar en la zona y en la República Argentina en general, es escasa (Podestá et. al, 2004; Raichijk et. al, 2005); por lo tanto, el seguimiento y estudio de este parámetro es importante especialmente para aplicaciones agrícolas, como por ejemplo dinámica de crecimiento de cultivos (Meinke et al., 1995), de contaminación atmosférica (dispersión de PM10, reacciones formadoras de NOx, SOx y formación de ozono; Seinfeld, 2001), además de tener interés primario en la cuantificación y seguimiento del recurso solar *per se*. No es de conocimiento de los autores que existan trabajos publicados de cuantificación directa de la radiación solar para la zona. Podestá et. al (2004), presentan resultados de la determinación de la radiación solar diaria a partir de las medidas de energía solar global de las estaciones del INTA-SMN, obtenidas durante los años 1981 y 1982 en Pergamino, a unos 65 km al sudoeste de San Nicolás. Sin embargo, existen otros trabajos relevantes que involucran a la zona a la hora de cuantificar el recurso solar y de determinar fórmulas predictivas, estos son los de Righini y Grossi Gallegos (2003), Raichik et al. (2005) y Raichijk y Grossi Gallegos (2007). Aunque los datos disponibles para este análisis pertenecen solamente al trienio 2006-2008, la posible discontinuidad en estas mediciones o la falta de disponibilidad de datos a futuro motivó la realización de un trabajo preliminar, utilizando los datos existentes para los años mencionados. No se dispone hasta el momento de datos más actuales.

La necesidad de conocer valores de radiación solar global y las dificultades generales para obtenerlos, han promovido la búsqueda de relaciones que permitan realizar predicciones a partir de determinados parámetros meteorológicos. La más común para la predicción de estos valores, es la bien conocida fórmula de Ångström (Ångström, 1924) posteriormente modificada por Prescott (Prescott, 1940), que relaciona la radiación solar diaria $Q$ de un sitio, generalmente en unidades del valor a tope de atmósfera o extraterrestre ($Qt$); o sea el cociente $Q/Qt$, el cual, por brindar una cuantificación de la radiación solar medida en tierra con respecto a la que teóricamente llega a un plano horizontal (respecto a la vertical local) por sobre la atmósfera, suele denominarse *transmisividad atmosférica o coeficiente de claridad K*. En el modelo predictivo de Ångström-Prescott (que originalmente fue elaborado para relacionar promedios mensuales de $K$, pero actualmente es usado en diversas escalas de tiempo; ver por ejemplo, Podestá et al., 2004), la variable independiente es la heliofanía relativa, que generalmente es un parámetro tanto o más escaso que los datos de radiación solar diaria. Por lo tanto, con la finalidad de poseer fórmulas más accesibles que permitan predecir valores de $Q$, se han ensayado otros métodos que tienen en cuenta variables meteorológicas de mayor disponibilidad o más fácilmente obtenibles. Particularmente, los métodos empíricos basados en la relación entre $K$ y la amplitud térmica diaria (la diferencia entre las temperaturas máxima y mínima del aire u otras combinaciones entre ellas), resultan muy convenientes, por ser la amplitud térmica una de las variables meteorológicas más ampliamente medidas (ver por ejemplo, Meinke et al., 1995; Meza y Varas, 2000; Raichijk et al., 2005; Bandyopadhyay et al., 2008).

El presente trabajo está básicamente inspirado en el de Raichijk et al. (2005), quienes han relacionado valores mensuales promedio de $Q$ en distintas localidades de Argentina, mediante un análisis de regresión lineal (método de Hargreaves o Hargreaves-Samani; citados por Bandyopadhyay et al., 2008) con los promedios mensuales de amplitud térmica diaria, de la siguiente forma:

$$<Q>/<Qt> = a + b\ <\Delta T>^{1/2} \qquad (1)$$

donde $a$ y $b$ son constantes reales; $<Q>$ es el promedio mensual de los valores calculados de $Q$; $<Qt>$ es el promedio mensual de los valores teóricos $Qt$ y $<\Delta T>^{1/2}$ es la raíz cuadrada del promedio mensual de la amplitud térmica diaria. El objetivo de este trabajo es estimar los coeficientes de la ecuación (1) con los datos obtenidos para San Nicolás con la finalidad de obtener relaciones preliminares que permitan estimar medias mensuales de radiación solar diaria en la zona, a partir de los promedios mensuales de amplitud térmica diaria.

## METODOLOGÍA

Los datos de irradiancia ($I$) se tomaron de una estación meteorológica DAVIS Vantage Pro2 que registra la radiación global incidente entre 400 y 1100 nm. Situada en la azotea de la Facultad Regional San Nicolás de la UTN, la estación está dedicada a reportar datos meteorológicos como apoyo de monitoreo ambiental desde finales del año 2005. La estación estuvo sujeta a calibraciones periódicas; por otra parte, los datos obtenidos han sido validados utilizando las facilidades del Centre Energétique et Procédés du Ecole des Mines de París (www.helioclim.net) (Geiger et al., 2002). Los datos de $I$ se obtuvieron para cada hora. La estación entrega el máximo ($I_M$) y el promedio ($I_m$) horario en Wm$^{-2}$. La radiación solar diaria $Q$ se calculó de acuerdo con la siguiente expresión:

$$Q = \int I\, dt \qquad (2)$$

la variable $t$ expresa el tiempo de horas de brillo del Sol por día (h/d); la integral se estimó utilizando $I_m$, y se extiende entre la hora de salida y de puesta del Sol. Para evitar posibles errores en las mediciones debido a la posición del Sol muy cercana del horizonte y para una compatibilidad con trabajos futuros, donde pueda aplicarse la expresión de Angström – Prescott, se ha decidido utilizar las medidas de irradiancia para las cuales $I_m > 120$ W m$^{-2}$ de acuerdo con las bien conocidas prescripciones de la World Meteorological Organization. Como medida de seguridad, se eliminaron la totalidad de las observaciones para aquellos días en que se detectó algún problema en la tomas de datos de $I$. De esta forma, el número total de valores calculados de $Q$ usados para el ajuste fue de 1011. Los valores $Qt$ se calcularon a partir del instante de salida y puesta solar para cada día, considerando la posición del Sol por sobre 5° de altura a fin de computar horas de irradiancia efectiva, siguiendo los procedimientos estándar (ver por ejemplo, Meza y Varas, 2000). Se ha tenido en cuenta además la corrección estacional (por excentricidad de la órbita terrestre) a la "constante solar" $I_0$, cuyo valor medio adoptado para los tres años de estudio fue de 1365,5 W m$^{-2}$ (Fröhlich, 2009): este valor se modula con la distancia Tierra-Sol ($r$) de tal forma que $I_0 = 1365,5$ W m$^{-2}$ /$(r/UA)^2$, donde UA es la unidad astronómica; los valores de $r$ se obtuvieron a partir de integraciones orbitales precisas (Cionco et al., 2012). En particular para el cálculo de la declinación solar ($\delta$) se ha usado:

$$\delta = -23{,}46° \cos\left(\frac{2\pi}{371{,}64} d + 0{,}23\right) \tag{3}$$

donde $d$ es un entero que representa el día del año (Cionco et al., 2007). Respecto a la amplitud térmica, se usaron los datos de temperatura para los mismos días considerados en la estimación de $Q$. Todos los cálculos se realizaron en ForTran 77 estándar en un entorno Linux, desarrollándose la casi totalidad de los programas utilizados.

**RESULTADOS**

La Fig. 1 muestra los valores calculados de $Q$ y los valores teóricos $Qt$ para cada día considerado. La evolvente superior de los puntos graficados, que es una estimación del coeficiente *K de días claros* (sin nubosidad), es aproximadamente 0,79 x $Qt$, lo cual está de buen acuerdo con lo reportado por Podestá et al. (2004) para Pergamino. La totalidad de los valores de $Q$, oscilan entre 0,3 y 36 MJ m$^{-2}$ d$^{-1}$. El cociente de los promedios mensuales $<Q>/<Qt>$ (36 valores en total, correspondiente a los tres años de estudio) y las raíces cuadradas de las amplitudes térmicas promedio se muestran en la Fig. 2. El resultado del ajuste por cuadrados mínimos arroja $a = 0{,}110$; $b = 0{,}181$ °C$^{-1}$, con una valor RMS residual de 0,066 MJ m$^{-2}$ d$^{-1}$ representando un RMS porcentual (RMS%) de 9,690 %.

El RMS residual se define como:

$$\text{RMS} = \sqrt[2]{\frac{\sum_{i=1}^{n}(Ko_i - Kc_i)^2}{n}} \tag{4}$$

donde $Ko_i$ son los cocientes $<Q>/<Qt>$ "observados", esto es, utilizados para el ajuste y calculados según lo indicado en la sección precedente; $Kc_i$ son los valores $<Q>/<Qt>$ obtenidos a partir del ajuste; $n$ es el número total de datos utilizados en el ajuste. El RMS% se obtiene dividiendo el RMS por el promedio de los $Ko_i$ utilizados en el ajuste y multiplicando por 100.

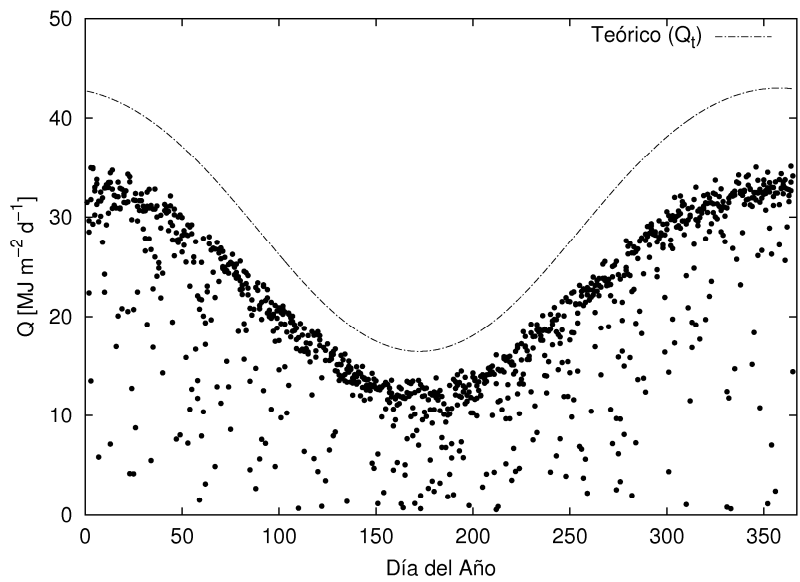

*Figura 1: Valores observados de radiación solar diaria utilizados para el ajuste (puntos) y valores teóricos por fuera de la atmósfera (línea punteada).*

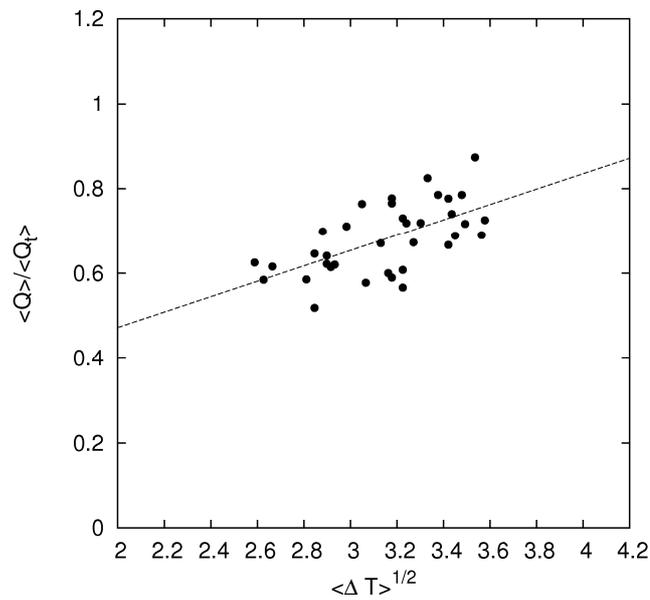

*Figura 2: Recta de regresión ajustada al cociente de los promedios mensuales de Q y Qt en función de la raíz cuadrada de la amplitud térmica mensual promedio. El coeficiente de correlación R es 0,848.*

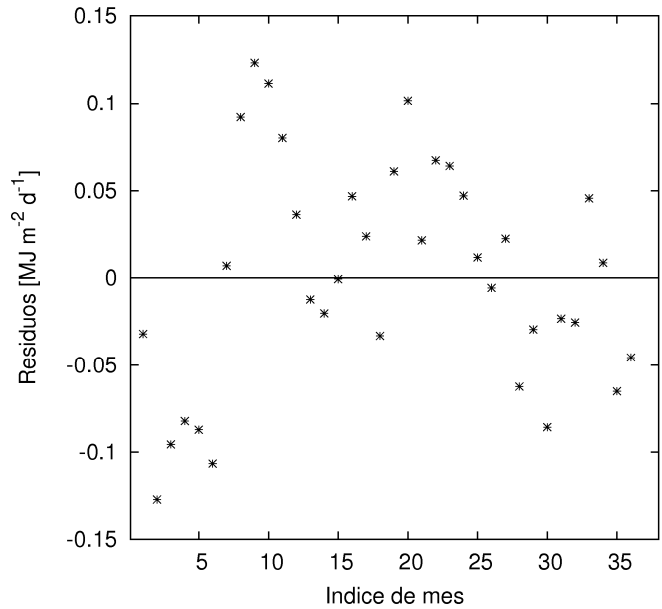

*Figura 3: Residuos post-ajuste en función del mes considerado. Se observan algunos efectos sistemáticos.*

De esta forma, la estimación de la media mensual de $Q$ para un mes determinado, se obtiene hallando los promedios mensuales de amplitud térmica y de $Qt$, introduciendo estos valores en la ecuación (1) con los coeficientes ajustados:

$$<Q> = <Qt> (0,110 + 0,181 \,°C^{-1} <\Delta T>^{1/2}). \tag{5}$$

La Fig. 3, muestra los residuos post-ajuste. Los valores absolutos de los residuos son siempre menores que 0,15 WJ m$^{-2}$ d$^{-1}$, manteniéndose bien por dentro del límite ± 3 RMS. Se observan algunos efectos sistemáticos: mientras que los primeros 12 valores (primer año) se distribuyen con bastante normalidad, los residuos correspondientes a los otros años (del 12 al 24 y del 25 al 36) muestran sistemáticamente mayor número de residuos positivos (segundo año) y negativos (tercer año). Sin embargo, la distribución de los residuos dentro del límite señalado, indica que los efectos sistemáticos no son importantes (no existen valores "outliers" significativos; Arley, 1950; Bevington, 1968).

**COMENTARIOS FINALES Y CONCLUSIONES**

En este trabajo se han determinado los coeficientes de un modelo lineal que relaciona valores mensuales promedio de $Q$ y la raíz cuadrada de medias mensuales correspondientes de amplitud térmica diaria, en base a una serie de tres años de datos de irradiancia solar, obtenida en la Facultad Regional San Nicolás de UTN. Los resultados indican que la utilización de la ecuación (1) para establecer dicha relación es muy razonable y su estudio en la zona debe profundizarse con más datos. Al respecto, es importante mencionar que se han ensayado otras relaciones posibles, distintas a la ecuación (1); por ejemplo, la relación de Bristow-Campbell (Bristow y Campbell, 1984), o el ajuste de un coeficiente distinto a ½ en la potencia de $<\Delta T>$, pero los resultados decididamente fueron de menor calidad estadística.

Por último, los resultados obtenidos están de buen acuerdo con lo reportado por Raichijk et al. (2005, tabla 1), obteniéndose valores concordantes con los de Paraná, Entre Ríos, a unos 150 km al norte de San Nicolás. El RMS % obtenido indica que el error esperado en la predicción de los valores de $Q$ también está dentro de lo estimado por esos autores. Los efectos sistemáticos presentes en los residuos deben estudiarse en función de un mayor número de observaciones.

**AGRADECIMIENTOS**